\documentclass[twoside]{BioRxiv}

\usepackage{graphicx}
\graphicspath{{figures/}}
\usepackage{epstopdf}
\usepackage{booktabs}
\usepackage{enumitem}
\usepackage{verbatim}
\usepackage{dblfloatfix}

\usepackage{tikz,lipsum,lmodern}
\usepackage[most]{tcolorbox}
\usepackage{array}

\newcommand\blfootnote[1]{%
 \begingroup
 \renewcommand\thefootnote{}\footnote{#1}%
 \addtocounter{footnote}{-1}%
 \endgroup
}

\begin{document}

\title{Measuring and modeling the motor system with machine learning}
\shorttitle{Measuring \& Modeling the Motor System}
\author[1,$\dagger$]{Sébastien B. Hausmann}
\author[1,$\dagger$]{Alessandro Marin Vargas}
\author[1,*]{Alexander Mathis}
\author[1,*]{Mackenzie W. Mathis}

\leadauthor{Hausmann, Marin Vargas et al.}
\affil[1]{EPFL, Swiss Federal Institute of Technology, Lausanne, Switzerland | $\dagger$co-first *co-senior} 

\maketitle

\begin{abstract}
The utility of machine learning in understanding the motor system is promising a revolution in how to collect, measure, and analyze data. The field of movement science already elegantly incorporates theory and engineering principles to guide experimental work, and in this review we discuss the growing use of machine learning: from pose estimation, kinematic analyses, dimensionality reduction, and closed-loop feedback, to its use in understanding neural correlates and untangling sensorimotor systems. We also give our perspective on new avenues where markerless motion capture combined with biomechanical modeling and neural networks could be a new platform for hypothesis-driven research. 

\end{abstract}
{\bf Highlights:}
\begin{enumerate}
\setlength\itemsep{-0.2em}
\item Deep learning-based tools allow for robust automation of movement capture and analysis

\item New approaches to modeling the sensorimotor system enable new hypotheses to be generated

\item These tools are poised to transform our ability to study the motor system

\end{enumerate}

\section*{Introduction}

\begin{figure*}
\begin{center}
\includegraphics[width=\textwidth]{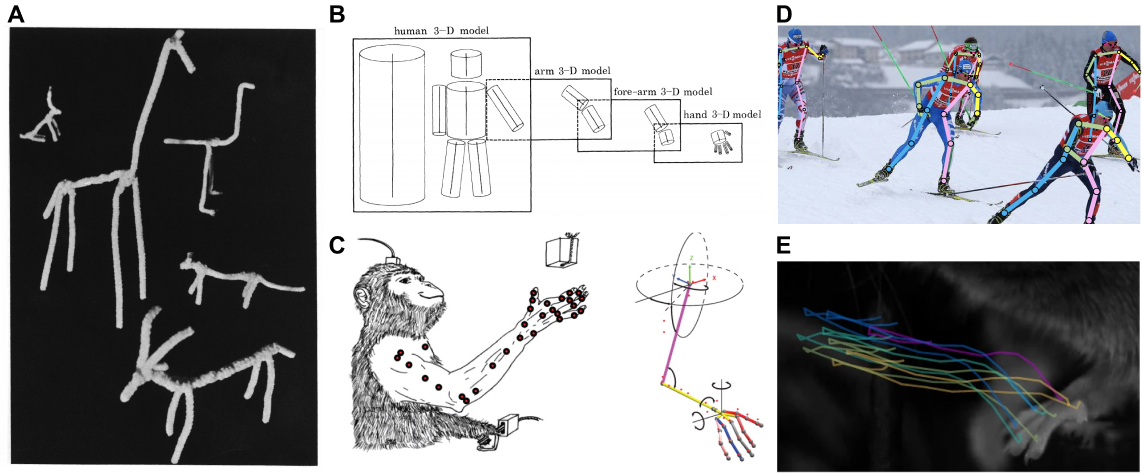}
\end{center}
\caption{{\bf Measuring movement:} Keypoint-based ``skeletons'' provide a reduced representation of posture. From Marr (A, B, adapted from~\citealp{marr1978representation}) to now, classical and modern computer vision has greatly impacted our ability to measure movement (C adapted from~\citealp{vargas2010decoding}, D adapted from~\citealp{wang2020deep}, and E adapted from~\citealp{mathis2018deeplabcut}).}
\label{fig:1}
\end{figure*}

Investigations of the motor system have greatly benefited from theory and technology. From neurons to muscles and whole-body kinematics, the study of one of the most complicated biological systems has a rich history of innovation. In their quest to understand the visual system, David Marr and H. Keith Nishihara described how static shape powerfully conveys meaning (information), and gave two examples of how ``skeletonized'' data and 3D shapes, such as a series of cylinders can easily represent animals and humans (Figure~\ref{fig:1}A,B; \citealp{marr1978representation}). These series of shapes evolving across time constitute actions, motion, and what we perceive as the primary output of the motor system. Now with deep learning, the ability to measure these postures has been massively facilitated. But the measurement of movement is not the sole realm where machine learning has increased our ability to understand the motor system. Modeling the motor system is also leveraging new machine learning tools. This opens up new avenues where neural, behavioral, and muscle-activity data can be measured, manipulated, and modeled with ever-increasing precision and scale. 

\blfootnote{\noindent * alexander.mathis@epfl.ch, mackenzie@post.harvard.edu}

\vspace{-20pt}
\section*{Measuring motor behavior} 

\justify Motor behavior remains the ultimate readout of internal intentions. It emerges from a complex hierarchy of motor control circuits in both the brain and spinal cord, with the ultimate output being muscles. Where these motor commands arise and how they are shaped and adapted over time still remains poorly understood. For instance, locomotion is produced by central pattern generators in the spinal cord and is influenced by goal-directed descending command neurons from the brainstem and higher motor areas to adopt specific gaits and speeds---the more complex and goal-directed the movement, the larger number of neural circuits recruited~\cite{Esposito2014BrainstemNM, bouvier2015descending}.
Of course, tremendous advances have already been made in untangling the functions of motor circuits since the times of Sherrington~\cite{sherrington1952integrative}, but new advances in machine learning for modeling and measuring the motor system will allow for better automation and precision, and new approaches to modeling, as we highlight below.
\medskip

In the past few years modern deep learning techniques have enabled scientists to closely investigate behavioral changes in an increasingly high-throughput and accurate fashion. Often eliminating hours of subjective and time-consuming human annotation of posture, ``pose estimation''---the geometric configuration of keypoints tracked across time---has become an attractive workhorse within the neuroscientific community~\cite{datta2019computational, Mathis2020DeepLT,pereira2020quantifying,von2021big}. While there are still computer vision challenges in both human and animal pose estimation, these new tools perform accurately within the domain they are trained on~\cite{mathis2021pretraining,koh2020wilds}.
Naturally, the question arises: now that pose estimation is accelerating the pace and ease at which behavioral data can be collected, how can these extracted poses help explain the underlying neural computations in the brain? 

\vspace{-4pt}
\subsection*{Meaningful data from pose estimation}

\justify Advances in pose estimation have opened new research possibilities. 
Software packages that tackle the needs of animal pose estimation with deep neural networks (DNNs) include DeepLabCut~\cite{mathis2018deeplabcut, nath2019deeplabcut}, LEAP~\cite{pereira2019fast}, DeepBehavior~\cite{arac2019deepbehavior}, DeepPoseKit~\cite{graving2019fast}, DeepFly3D~\cite{gunel2019deepfly3d}, and DeepGraphPose~\cite{wu2020deep}.
These packages provide customized tracking for the detailed study of behavior, and come with their own pitfalls and perspectives~\cite{mathis2020primer}. Pose estimation was reviewed elsewhere~\cite{mathis2020primer, Mathis2020DeepLT,pereira2020quantifying, von2021big}, which is why we focus on how such data can be used to understand the motor system.
\medskip

Several groups have delved into the challenge of deriving additional metrics from deep learning-based pose estimation data. There are (at least) two common paths: (1) derive kinematic variables, (2) derive semantically-meaningful behavioral actions. Both have unique challenges and potential uses. 

\subsection*{Kinematic analysis}

\justify Whether marker-based or marker-free, tracked keypoints can be viewed as the raw building blocks of joint coordinates and subsequent measurements. Namely, key points can form the basis of movement analysis. Kinematics---the analysis of motion, and in biomechanics typically the study of position, velocity, and acceleration of joints---is essential to describe the motion of systems. 
\medskip

How has kinematic data been used to understand the motor system? In elegant work, Vargas-Irwin and colleagues demonstrated how highly detailed pose estimation data can be used to uncover neural representations in motor cortex during skilled reaching. Using a marker-based approach, they showed that a small number of primary motor cortex (M1) neurons can be used to accurately decode both proximal and distal joint locations during reaching and grasping~\cite{vargas2010decoding} (Figure~\ref{fig:1}C). Others have powerfully used marker-based tracking to quantify recovery in spinal cord injuries in mice, rats and macaques~\cite{courtine2008recovery,von2016neurorobotic, van2012restoring}.
\medskip

Markerless pose estimation paired with kinematic analysis is now being used for a broad range of applications. Human pose estimation tools, such as state-of-the-art (on COCO) HRNet~\citep{wang2020deep} (Figure~\ref{fig:1}D) or DeepLabCut (Figure~\ref{fig:1}E), have been used in applications such as sports biomechanics~\cite{kidzinski2020deep,white2020relationship}, locomotion (Figure~\ref{fig:2}) and clinical trials\footnote{\url{https://clinicaltrials.gov/ct2/show/NCT04074772}}. For example, Williams et al. recently showed that finger tap bradykinesia could be objectively measured in people with Parkinson's disease~\cite{williams2020discerning}. They used deep learning-based pose estimation methods and smartphone videos of finger tapping kinematics (speed, amplitude and rhythm) and found correlations with clinical ratings made by multiple neurologists. With a broader adoption of markerless approaches in both the biomechanical and neuroscience community, we foresee a growing use of machine learning for developing clinically-relevant biomarkers, and automating the process of clinical scoring of disease states (i.e., automating scoring Parkinson's disease).

\subsection*{Reducing dimensionality to derive actions}

\justify Analyzing and interpreting behavior can be challenging as data are complex and high-dimensional. While pose estimation already reduces the dimensionality of the problem significantly, there are other processing steps that can be used to transform video into ``behavior actions''. Specifically, dimensionality reduction methods can transform the data into a low-dimensional space, enabling a better understanding and/or visualization of the initial data.

\medskip

\begin{figure*}[hb]
\begin{center}
\includegraphics[width=.95\textwidth]{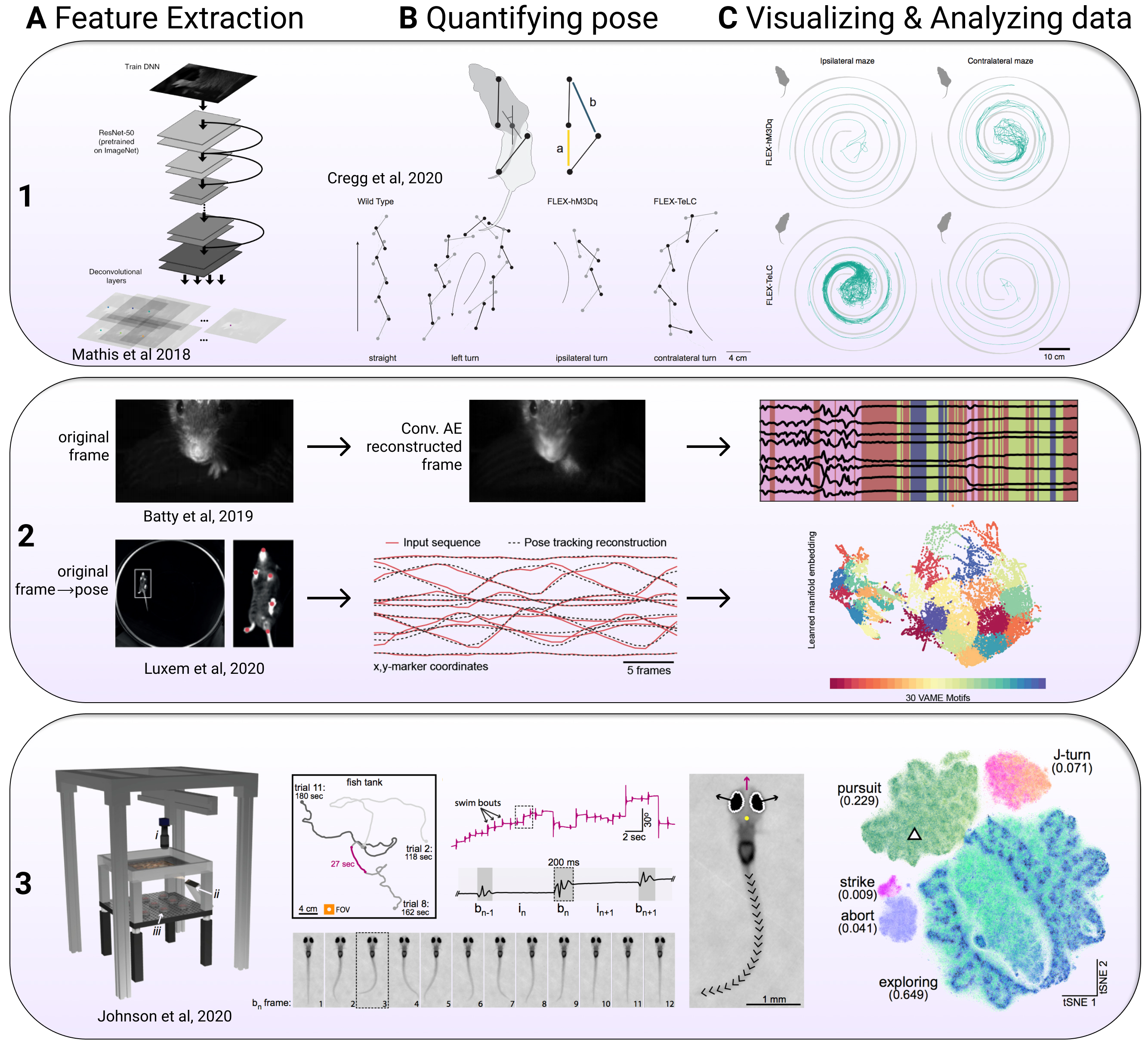}
\end{center}
\caption{{\bf From Video to Behavioral Metrics:} How deep learning tools are shaping new studies in neural control of locomotion, analyzing natural behavior, and kinematic studies. The general workflow consists in (A) extracting features (pose estimation), (B) quantifying pose and (C) visualizing and analyzing data. (1) Limb tracking to analyze gait patterns in mice whose locomotion has been altered by brainstem optical stimulation (adapted from~\citealp{mathis2018deeplabcut, cregg2020brainstem}). (2) Autoencoders: Top row: The use of a convolutional autoencoder from video to inferred latents and states, adopted from~\citealp{batty2019behavenet}. Bottom row: Architecture of VAME (adapted from~\citealp{luxem2020identifying}) constituted by an encoder (input is key points), which learns a representation of the hidden states. The original sequence can be reconstructed by a decoder or the evolution of the time series can be predicted. UMAP embedding of the pose estimation. (3) Behavioral data acquisition enabling precise tracking of zebrafish larvae and behavioral clustering of swimming bouts (adapted from~\citealp{johnson2020probabilistic}).}
\label{fig:2}
\end{figure*}

Dimensionality reduction tools (e.g., PCA, t-SNE~\cite{maaten2008visualizing}, UMAP~\cite{mcinnes2018umap}) are commonly used methods to cluster data in an unsupervised manner. Principal Component Analysis (PCA) aims to find the components that maximize the variance in the data and the principal components rely on orthogonal linear transformations, and has been important for estimating the dimensionality of movements, for example~\cite{vargas2010decoding,yan2020unexpected}. t-distributed Stochastic Neighbor Embedding (t-SNE) is suitable for visualizing non-linear data and seeks to project data into a lower dimensional space such that the clustering in the high dimensional space is preserved. However, it \textit{typically} does not preserve the global data structure (i.e., only within-cluster distances are meaningful; but cf.~\citealp{kobak2019art}). In conjunction with an impressive system to track hunting zebrafish, t-SNE was used to visualize the hunting states (Figure~\ref{fig:2}). Lastly, Uniform Manifold Approximation and Projection for Dimension Reduction (UMAP) typically preserves the data’s global structure (cf.~\citealp{kobak2021initialization}). Many tools allow for seamless passing of pose estimation output to perform such clustering, such as MotionMapper~\cite{Berman2014} and B-SOiD~\cite{hsu2020b}.
\medskip

The utility of unsupervised clustering for measuring behavior is well illustrated by Bala et al.~\citealp{bala2020openmonkeystudio}. They explored motor behavior in freely moving macaques in large unconstrained environments in the laboratory setting. In combination with 62 cameras---a remarkable feat on its own---OpenMonkeyStudio uses 13 body landmarks and trains a generalizable view-invariant 2D pose detector to then compute 3D postures via triangulation. Coherent clusters extracted with UMAP correlated with semantic actions such as sitting, climbing, and climbing upside down in monkeys~\cite{bala2020openmonkeystudio}. Importantly, in this case, a 3D pose estimation compared to 2D pose estimation yielded more meaningful clusters. Together with the 3D location of a second macaque, social interactions were derived from co-occurrence of actions. 
\medskip

Dimensionality reduction can also be applied to kinematic features (such as joint angles). For example, DeAngelis et al. used markerless pose estimation to extract gait features from freely moving hexapods~\cite{deangelis2019manifold}. Then, UMAP was applied to generate a low-dimensional embedding of the tracked limbs while preserving the local and global topology of the high-dimensional data. UMAP revealed a vase-like manifold parametrized by the coordination patterns of the limbs. In conjunction with modulation by means of optogenetic or visual perturbations, different gaits and kinematic modalities were evoked and directly interpretable in the UMAP embedding. In a similar approach, using the clustering algorithm \textit{clusterdv}~\cite{marques2019clusterdv}, different types of swimming bouts were identified from zebrafish larvae movements~\cite{marques2018structure}. Similar methods enabled precise visualization of hunting strategies that were dependent on visual cues~\cite{mearns2020deconstructing}.
\medskip

Autoencoders are powerful tools for nonlinear dimensionality reduction via generative models~\cite{kingma2013auto}. They have found wide use in sequence modeling for language, text, and music~\cite{Goodfellow-et-al-2016, Blei2016VariationalIA}. They can also be used for unsupervised discovery of behavioral motifs by using manually selected features (i.e., the keypoints defined for pose estimation) as input, or by inputting video (Figure~\ref{fig:2}. For instance, BehaveNet uses video directly to model the latent dynamics~\cite{batty2019behavenet}. Variational Animal Motion Embedding~\cite{luxem2020identifying} (Figure~\ref{fig:2}) used keypoints as the input to a recurrent variational auto encoder to learn a complex distribution of the data.
Variational autoencoder stochastic neighbor embedding (VAE-SNE) combines a VAE with t-SNE to compress high dimensionality data and automatically learn a distribution of clusters within the data~\cite{graving2020vae}.
\medskip

In addition to unsupervised analysis, supervised methods are powerful ways to use the outputs of pose estimation to derive semantically meaningful actions~\cite{Pedregosa2011, Nilsson2020SimpleBA, Sturman2020DeepLB, segalin2020mouse}. For instance, the Mouse Action Recognition System (MARS), a pipeline for pose estimation and behavior quantification in pairs of freely behaving (differently colored) mice, works in conjunction with BENTO, which allows users to annotate and analyze behavior states, pose estimates, and neural recording data simultaneously through a graphical interface~\cite{segalin2020mouse}. This framework enabled the authors to use supervised machine learning to identify a subset of 28 neurons whose activity was modulated by mounting behavior. These methods build on open source tools, which provided many supervised and unsupervised learning tools in a highly customize way~\cite{Pedregosa2011}.
\medskip

Another approach is to use video directly, instead of applying pose estimation first~\cite{Berman2014, wiltschko2015mapping,batty2019behavenet}. Here, the pixels themselves can be used for action recognition, which can be highly useful when the behaviors of interest do not involve kinematics, such as blushing in humans, or freezing in mice.
In computer science the rise of deep learning has gone hand-in-hand with the development of larger datasets. For example, pre-training models on the Kinetic Human Action Video dataset drastically improves action recognition performance~\cite{carreira2017quo}. Others pushed video recognition further by using a \textit{multi-fiber} architecture, where sparse connections are introduced inside each residual block, thereby reducing computations~\cite{chen2018multi}. 
Several groups have leveraged this approach for facial expression~\cite{andresen2020towards,dolensek2020facial}, pharmacological behavior-modulation~\cite{wiltschko2020revealing} or for measuring posture and behavior representations~\cite{brattoli2017lstm, bohnslav2020deepethogram}. 

\subsection*{Closed-loop feedback based on behavior}

\justify Closed-loop feedback based on behavioral measurements can be informative for causal testing of learning algorithms (such as reinforcement learning) and for probing the causal role of neural circuits in behavior~\cite{buckley2018theory, nourizonoz2020etholoop}. Recent efforts to translate offline pose estimation and analysis to real time have enabled new systems, such as in EthoLoop~\cite{nourizonoz2020etholoop} and DeepLabCut-live!~\cite{kane2020real}.
EthoLoop is a multi-camera, closed-loop tracking system using a two-step process: a marker-based camera tracking system, followed by DeepLabCut-based pose estimation analysis. This system is capable of providing close-up views and analyses of the ethology of tracked freely moving primates. The closed-loop aspect enabled real-time wireless neuronal recordings and optical stimulation of individuals striking specific poses. We expect that real-time (and predictive) low-latency pose estimation for closed-loop systems will certainly play a crucial role in the years to come~\cite{kane2020real, forys2020real,Schweihoff2019DeepLabStreamCT}. These tools are bound to include more customized behavior-dependent real-time feedback options. While the delays can be minimal for such computations, time-delayed (hardware) systems are not real-time controllable. Thus, to instantly provide feedback we added a forward prediction mode to DeepLabCut-Live!~\cite{kane2020real}, which we believe will be crucial as these tools grow in complexity. Overall, the above discussed methods are highly effective in disentangling behavioral data, finding patterns, and capturing actions within behavioral data.

\subsection*{Neural correlates of behavior}

\justify At the heart of neuroscience research is the goal to causally understand how neurons (from synapses to networks) relate to behavior. With the rise of new tools to measure behavior, there is a homecoming to the quest of relating movement to neural activity. As described above, both kinematic features or lower dimensional embeddings of behavior can be used to regress against neural activity, or using real-time feedback, to causally probe their relationship to actions.
\medskip

In recent years, many groups have utilized such tools to uncover new knowledge of the motor system. It remains debated what the role of motor cortex is, yet new tools are enabling careful and precise studies of the system in goal-directed actions. For example, Ebina et al. used markerless tracking and optogenetics to show that in marmosets stimulation with varying spatial or temporal patterns in motor cortex could elicit simple or even direction-specific forelimb movements~\cite{ebina2019arm}. Sauerbrei et al. also used markerless tracking to show that motor cortex inactivation halted movements, but this was a result of disrupted inputs (from thalamus), thus revealing that multiple interacting brain regions were responsible for dexterous limb control~\cite{sauerbrei2020cortical}. Moreover, others have used these tools to show that brainstem neurons highly correlate and causally drive locomotor behaviors~\cite{van2020freely, cregg2020brainstem}.
\medskip

Furthermore, it is becoming increasingly recognized that brain-wide (or minimally, cortex-wide) neural correlates of movement are ubiquitous in animals performing both spontaneous or goal-directed actions~\cite{Stringereaav7893, Musall_NN, Mathis2019ANS}. Stringer et al.~\cite{Stringereaav7893} showed spontaneous movements constituted much of the neural variance (``noise'') in visual cortex, and Musall, Kaufmann et al.~\cite{Musall_NN} report similar findings across many brain regions. It was also previously shown that neuronal activity in the posterior parietal cortex and the pre-motor cortex (M2) of rats accurately correlate with animal’s head posture~\cite{mimica2018efficient}, and of course even in sensory areas such as visual cortex encode movement~\cite{Niell2010ModulationOV, Leinweber2017ASC, Stringereaav7893}. 
\medskip

Relating neural activity to not only kinematic or behavioral features, but to muscle output is also highly important for correlating neurons to movement. In a recent study, human neonatal motoneuron activity was characterized using noninvasive neural interface and joint tracking. Using markerless pose estimation, they found that fast leg movements in neonates are mediated by high motoneuron synchronization, and not simply due to an increase in discharge rate as previously observed in adults~\cite{del2020spinal,vecchio2019neonates}. 
Another example of combining EMG (electromyography) and motion capture, Herent et al. used peripheral recordings via a diaphragmatic EMG to reveal an absence of synchronization (e.g., temporal correlation) of breaths to strides in mice at various displacement speeds during locomotion~\cite{herent2020independent}. This rich data is crucial for understanding the neural code, and it is shaping efforts to model the system at an even finer resolution.
\medskip

Being able to not just capture movement, but model it has a rich history in movement neuroscience~\cite{Scott2004OptimalFC, todorov2002optimal, franklin2011computational}. In the next sections we will discuss how machine learning has also influenced modeling the motor system.

\vspace{-5pt}
\section*{Neural networks as sensorimotor system models}

\justify The brain excels at orchestrating adaptive animal behavior, striving for robustness across environments~\cite{sherrington1952integrative, franklin2011computational}. Thereby the brain, hidden in the skull, takes advantage of multiple sensory streams in order to act as a closed-loop controller of the body. How can we elucidate the function of this complex system? 
The solution to this problem is not trivial due to multiple challenges such as high-dimensionality, redundancy, noise, uncertainty, non-linearity and non-stationarity of the system~\cite{franklin2011computational}. Since DNNs excel at learning complex input-output mappings~\cite{poggio2020theoretical,yang2020artificial}, they are well positioned for modeling motor, sensory and also sensorimotor circuits. Moreover, unlike in the biological brain, DNNs are fully observable such that one can easily ``record'' from all the neurons in the system and measure the connectome. Therefore, in the next sections we discuss how modeling the motor, sensory and combine sensorimotor systems may lead to new principles of motor control.

\vspace{-5pt}

\subsection*{Modeling the motor system}

\justify How do neural circuits produce adaptive behavior, and how can deep learning help model this system? Several groups have modeled the motor system with neural networks to investigate how, for example, task cues can be transformed into rich temporal sequential patterns that are necessary for creating behavior~\cite{churchland2012neural, vyas2020computation}. In particular, recurrent neural networks (RNNs), whose activity is dependent on their past activity (memory), have been used to study the motor system, as they can produce complex dynamics~\cite{yang2020artificial,poggio2020theoretical}.
\medskip

In a highly influential study, Sussillo and colleagues showed that RNNs can learn to reproduce complex patterns of muscle activities recorded during a primate reaching task. By modifying the network’s characteristics, such as forcing the dynamics to be smooth, the natural dynamics that emerged from the network closely resembled the one observed in the primates’ primary motor cortex (M1)~\cite{sussillo2015neural}. In a similar way, RNNs can also be used to understand why a brain area shows a specific feature. For instance, M1 has low-tangled population dynamics when primates perform a cycling movement task using a hand-pedal, unlike muscle activity and sensory feedback~\cite{russo2018motor}. Building a network model which is trained on the same task, not only can replicate the same dynamics but it also unveils noise robustness as a possible underlying principle of the observed dynamics. Moreover, RNNs can be used to test various hypotheses about how the brain drives adaptive behavior by comparing neural or behavioral activity to the ``neural'' units. For instance, it has been investigated how prior beliefs are integrated into the neural dynamics of the network~\cite{sohn2019bayesian}, how temporal flexibility is connected to a network's nonlinearities~\cite{wang2018flexible}, and how robust trajectory shifting allows the translation between sequential categorical decisions~\cite{chaisangmongkon2017computing}. 
\medskip

Importantly for motor control, RNNs have also been used to study how sequential, independent movements, or tasks, are generated~\cite{zimnik2021independent}. Here, the authors also tackled the problem of cross task interference, which the RNN overcame by utilizing orthogonal subspaces~\cite{zimnik2021independent}. In related work, a novel learning rule that aimed to conserve network dynamics within subspaces (defined by activity of previously learned tasks) allowed for more robust learning of multiple tasks with a RNN~\cite{duncker2020organizing}. 
\medskip

RNNs are also capable of re-producing complex spatio-temporal behavior, such as speech~\cite{anumanchipalli2019speech, makin2020machine}. This was achieved using a two-stage decoder with bidirectional, long short-term memory networks. First, the articulatory features were decoded by learning a mapping between sequences of neural activity (high-gamma amplitude envelope and low frequency component from EcoG recordings) and 33 articulatory kinematic features. Second, the kinematic representation is mapped to 32 acoustic features. Importantly, because subjects share the kinematic representation, the first stage of the decoder could be transferred to different participants, which requires less calibration data~\cite{makin2020machine}.

\begin{figure*}[t]
\begin{center}
\includegraphics[width=.96\textwidth]{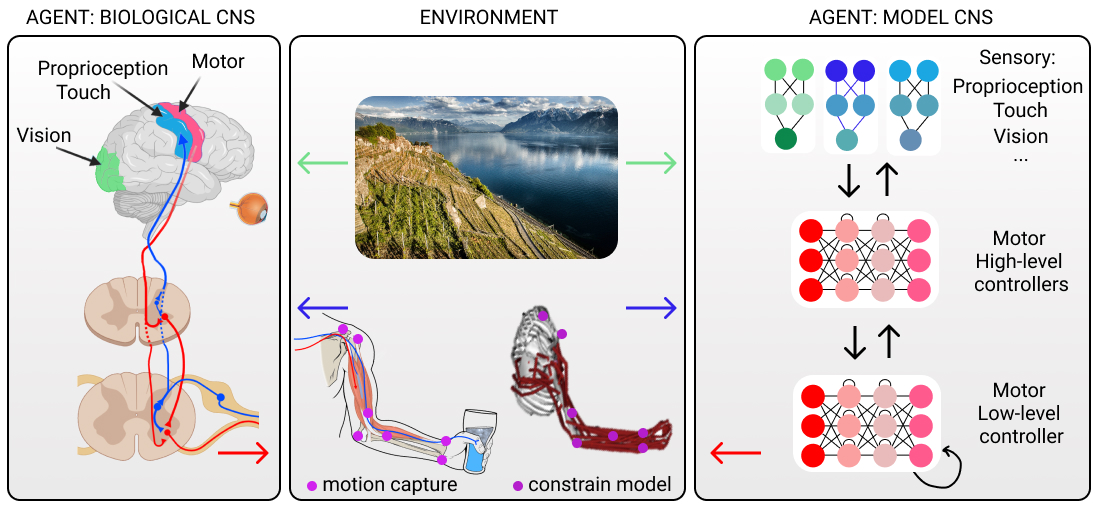}
\end{center}
\caption{{\bf Modeling the sensorimotor system}. 
The biological agent's central nervous system (CNS, left panel) receives sensory input such as vision, proprioception, and touch from the environment (middle panel). The sensorimotor system can be modeled using hierarchical, artificial neural networks (right panel), such as feed-forward networks (blue and green), and recurrent networks (red) to represent the sensory and motor systems (respectively). The biological agent and environment provide crucial model constraints (derived from fine-grained behavioral measurements, biochemical and anatomical constraints, etc.). For instance, motion capture and modeling of the plant (body) in the environment (i.e., here the arm) can be used to constrain motor outputs. The model CNS can be used to make predictions about the biological system, and used to explain biological data.
Anatomical drawings on the left created with BioRender.com; OpenSim arm adapted from~\cite{sandbrink2020task}.}
\label{fig:3}
\end{figure*}

\vspace{-5pt}
\subsection*{Modeling sensory systems}

\justify Sensory inputs, which are delayed and live in a different coordinates to the motor system, need to be integrated for adaptive motor control~\cite{todorov2002optimal, franklin2011computational}. Deep learning based models of sensory systems have several advantages: responses for arbitrary stimuli can be computed, their parts can be mapped to brain regions, and they have high predictability~\cite{yamins2014performance, yamins2016using, kell2019deep}.
\medskip

In general, there are two main approaches used for generating models of sensory systems: (i) data-driven models and (ii) task-driven models. Data-driven approaches use encoding models that are trained to predict neural activity from stimuli. As such, using non-deep-learning-based encoding models is a classical approach~\cite{gerven2017primer} where neural responses are approximated using tuning curves~\cite{prud1994proprioceptive}. Since stimulus features can be highly complex, DNNs can be especially useful in learning the stimulus-response mappings~\cite{yamins2016using, benjamin2018modern, cadena2019deep, kell2019deep, Walker2019InceptionLD, Bashivan2019NeuralPC}. Indeed, DNNs can outperform standard methods, such as the generalized linear model (GLM), in predicting the neural activity of the somatosensory and motor cortex~\cite{benjamin2018modern}.
Yet, to fit complex deep learning models, one typically needs a lot of data, a limitation that can be overcome by the task-driven approach. 
\medskip 

Task-driven modeling builds on the hypothesis that sensory systems have evolved to solve relevant ecological tasks~\cite{simoncelli2001natural, geisler2011contributions}: if artificial systems are trained to solve the same complex tasks that animals face, they might converge towards representations observed in biological systems~\cite{yamins2016using, kell2019deep,richards2019deep}. Therefore, one can (potentially) learn large-scale models for limited neural datasets by taking advantage of transfer learning (e.g., train on ImageNet~\cite{russakovsky2015imagenet} to obtain better models of the visual (ventral) pathway). Second, the choice of tasks allows researchers to test diverse hypotheses such as the highly successful hypothesis that the ventral pathway in primates is optimized to solve object recognition~\cite{yamins2014performance, yamins2016using,schrimpf2020integrative}. Specifically, hierarchical DNNs trained to solve an object recognition task can explain a large fraction of variance in neural data of different brain areas along the visual pathway~\cite{schrimpf2020integrative} and auditory pathway~\cite{kell2018task}. Crucially, for audition and vision, large scale datasets of relevant stimuli are readily available. How can this work be expanded to senses of importance to the motor system, such as touch and proprioception, where delivering relevant touch-only or proprioception-only information is more challenging (if not impossible)?
\medskip
 
One possible way to overcome this issue consists of simulating touch and proprioceptive inputs using biophysical models. For instance, a physically-realistic model of a mouse's whisker array has been used to develop a synthetic dataset of whisker (touch) sweeps across different 3D objects~\cite{zhuang2017toward}. A muscle-spindle firing rate dataset has been generated from 3D movements based on a musculoskeletal model of a human arm~\cite{sandbrink2020task}. 
In this way, DNNs trained to perform object or character recognition suggested putative ways spatial and temporal information can be integrated across the hierarchy~\cite{zhuang2017toward} and that network's architecture might play a main role in shaping its kinematic tuning properties~\cite{sandbrink2020task}. Both of these studies propose perception-based tasks, which provide baseline models, yet we envision the task space, complexity of the biophysical models, and the plant modelled will increase in future studies.

\subsection*{Modeling the sensorimotor system}

\justify Although modeling the motor or sensory systems alone is of great importance, a crucial body of work stems from combining models of sensory, motor, and task-relevant signals. Deep reinforcement learning (DRL) is a powerful policy optimization method to train models of sensorimotor control~\cite{botvinick2020deep}.
\medskip

For instance, DRL was used to solve a navigation task in zebrafish, and representations in units of the network resembled those observed in the brain~\cite{haesemeyer2019convergent}. Not only did the optimized network have units that correlated with temperature and behavioral states, but it also predicted an additional functional class of neurons which was not previously identified with calcium imaging alone~\cite{haesemeyer2019convergent}.
DNNs trained to achieve chemotaxis behavior of \textit{c. elegans} with DRL provided insights into neuropeptide modulation and circuits~\cite{kim2020deep}. In an artificial agent trained to perform vector-based navigation, grid cells emerged as well as coding properties~\cite{banino2018vector} that had been observed in rodents~\cite{stensola2012entorhinal}, and predicted from theory~\cite{stemmler2015connecting}. Motor control has been studied using a virtual rodent trained with DRL to solve a few behavioral tasks. A closer look into the learned neural representations delineates a task-invariant and a task-specific class, which belong to the low- and high-level controller, respectively~\cite{merel2019deep}. 
\medskip

Moreover, DRL has been utilized to learn control policies for complex movements, such as locomotion in challenging environments~\cite{heess2017emergence,peng2016terrain, peng2017deeploco} or complex dexterous hand manipulation~\cite{jain2019learning}. Eventually, the policy learned in simulation environments can be transferred to a real-world scenario achieving rather remarkable robustness in terrains not encountered during training~\cite{lee2020learning} and evolving human-like grasping and object interaction behaviors without relying on any human demonstrations~\cite{akkaya2019solving,andrychowicz2020learning}. Lastly, real-world sensory feedback and measurements during natural tasks can also be used to constrain models and define goals. For example, reference motion data can be used to train musculoskeletal model using DRL to reproduce human locomotion behaviors~\cite{zhou2019efficient, song2020deep}.
\medskip

An alternative approach to modeling the sensorimotor system in neural networks is via engineering or theoretical principles. 
Neural network models have an increased capacity to model highly complex integrated systems, as evidenced by advances, like Spaun, a spiking, multi-million-neuron-network model that can perform many cognitive tasks~\cite{Eliasmith2012ALM, DeWolf2016ASN, Choo2018Spaun2E}. Additionally, optimal feedback control (OFC) theory is a powerful normative principle for deriving models, that predicts many behavioral phenomena~\cite{todorov2002optimal, Scott2004OptimalFC}. OFC translated into DNNs accurately predicted neural coding properties in the motor cortex with biomechanical arm models and peripheral feedback~\cite{lillicrap2013preference}. Recent experimental and modeling work has also begun to link different cortical regions to their functional roles in an OFC theory framework~\cite{takei2020causal,mathis2017}. We believe that future work will combine neural recordings and perturbations to continue to test hypotheses generated by DNN-based OFC, and other DNN-based system-wide models, which are constrained by rich behavior (Figure~\ref{fig:3}).
\medskip

Towards such hybrid models, Michaels et al. have developed an exceptional model which leverages visual feedback to produce grasping movements. This model combines a CNN to extract visual features and three RNNs (modules) to control a musculosketal human arm. Amongst different architectures, the one with sparse inter-module connectivity was the best in explaining the brain-related neural activity thereby revealing possible anatomical principles of cortical circuits. Moreover, behavioral deficits, observed in previous lesion studies of the corresponding cortical areas, could be predicted by silencing specific modules. Interestingly, slight deficits occurred in regularized networks (i.e., penalty on high firing rates) when inputs from the intermediary module to the output module were lesioned, whereas behavior was completely disrupted in non-regularized networks. This observation suggests that the minimization of the firing rate could be a potential organizational principle of cortical circuits and that M1 might autonomously generate movements~\cite{michaels2020goal}. However, this contrasts recent evidence that continuous input from the thalamus (in mice) is necessary to perform movements~\cite{sauerbrei2020cortical}. 
\medskip

DNNs can also be trained in a self-supervised way based on rich data sets~\cite{orhan2020self}, which is a highly attractive platform for studying sensorimotor learning. Sullivan et al. showed that DNNs can learn high-level visual representations when trained on the same naturalistic visual inputs that babies receive using head-mounted cameras~\cite{sullivan2020saycam}. These learned representations are invariant to natural transformations and support generalization to unseen categories with few labelled examples~\cite{orhan2020self}. An important future direction for studying the motor system will be to not only focus on representational similarity, but also comparing the learning rules used by both biological systems and machine models. 
\medskip

Lastly, DNNs for sensorimotor control are also popular in robotics. Not only can robots serve as great testing grounds for control policies~\cite{Ebert2018VisualFM, Pearson2011BiomimeticVS}, but they can also reveal limitations when going from ``simulation to reality'', such as challenges related to robustness~\cite{Heaven2019WhyDA, Hendrycks2019NaturalAE}.

\section*{Outlook}

In the past few years, neuroscience has tremendously benefited from advances in machine learning. Behavioral analysis got more accurate while also being much less time consuming. This has already revealed novel aspects of behavior, but we are just at the dawn of these developments. Furthermore, advances in deep learning shaped the way biological systems can be modeled. We believe that in the future these two aspects will become increasingly intertwined. Namely, the unreasonable effectiveness of data suggests that powerful models---which further approximate the neural code---can be trained with large-scale behavioral measurements that are now possible. Concretely, with new tools to measure behavior one can constrain biologically plausible agents (such as OpenSim~\cite{delp2007opensim} or mujoco models~\cite{todorov2012mujoco}) to generate the behavior in an artificial setting. This simulation can then be used to generate data that might not otherwise be available: i.e., muscle activity from the whole arm or body in parallel with a visual input. These data could be used to develop hierarchical DNN models of the system (as we illustrate in Figure~\ref{fig:3}). These artificial network models, constrained with complex tasks, will provide researchers with sophisticated tools for testing hypotheses and gaining insight about the mechanisms the brain might use to generate behavior.

\vspace{10pt}
\begin{flushleft}
\textbf{Acknowledgments:} 
\end{flushleft}
The authors declare no conflicts of interest. We thank members of the Mathis Lab, Mathis Group and Travis DeWolf for comments. MWM is the Bertarelli Foundation Chair of Integrative Neuroscience.

\section*{References}

\end{document}